\documentstyle[11pt,fleqn]{article}
\def\titolo{\par\bigskip\begin{center}\bf\LARGE}
\def\endtitolo{\end{center}\par\bigskip\par\rm\normalsize}
\def\instit{\begin{center}\large}
\def\endinstit{\end{center}\rm\normalsize}
\def\references{\begin{thebibliography}{10}}
\def\endreferences{\end{thebibliography}\end{document}}

\newcommand{\btit}{\begin{titolo}}
\newcommand{\etit}{\end{titolo}}

\renewcommand{\author}[1]{\begin{center}\Large #1\end{center}}
\renewcommand{\date}[1]{\par\bigskip\par\sl\hfill #1\par\medskip\par}

\newcommand{\pacs}[1]{\smallskip\noindent{\sl PACS number(s):
                       \hspace{0.3cm}#1}\par\bigskip}
\newcommand{\babs}{\hrule\par\begin{description}\item{Abstract: }\it}
\newcommand{\eabs}{\par\end{description}\hrule\par\medskip\rm}
\newcommand{\ack}[1]{\par\section*{Acknowledgments} #1}
\newcommand{\ca}[1]{{\cal #1}}         
\newcommand{\nn}{\nonumber}            
\newcommand{\beq}{\begin{eqnarray}}    
\newcommand{\eeq}{\end{eqnarray}}      
\newcommand{\beqn}{\begin{eqnarray}}   
\newcommand{\eeqn}{\end{eqnarray}}     
\newcommand{\at}{\left(}               
\newcommand{\aq}{\left[}               
\newcommand{\ct}{\right)}              
\newcommand{\cq}{\right]}              
\newcommand{\ii}{\infty}                         
\newcommand{\fr}[2]{\mbox{$\frac{#1}{#2}$}}      
\newcommand{\Tr}{\,\mbox{Tr}\,}                  
\newcommand{\Res}{\,\mbox{Res}\,}                
\renewcommand{\Re}{\,\mbox{Re}\,}                
\newcommand{\lap}{\Delta}                        
\newcommand{\al}{\alpha}
\newcommand{\be}{\beta}

\newcommand{\ze}{\zeta}

\newcommand{\Ga}{\Gamma}

\begin{document}

\begin{center}

{\large \bf  HIGH TEMPERATURE EXPANSION OF STRING FREE ENERGY \\
             IN HYPERBOLIC SPACE}

\vspace{4mm}

{\sc A.A. Bytsenko} \\ {\it Department of Theoretical Physics,
State Technical University, \\ St Petersburg 195251, Russia} \\
{\sc S.D. Odintsov}\footnote{E-mail address: odintsov@ecm.ub.es.}
 \footnote{On leave from Tomsk Pedagogical Institute,Tomsk,Russia}\\
{\it Department E.C.M., Faculty of Physics, University of
Barcelona, \\
Diagonal 647, 08028 Barcelona, Spain} \\
{\sc S. Zerbini} \footnote{E-mail address: zerbini@science.unitn.it}\\
{\it Department of Physics, University of Trento, 38050 Povo,
Italy}    \\ and
{\it I.N.F.N., Gruppo Collegato di Trento}


\end{center}

\vspace{5mm}

\begin{abstract}
The high temperature behaviour of the open  bosonic  string  free
energy in  the  space $S^1 \otimes H^N$
with vanishingly  small  curvature  is investigated. The leading term of the
high temperature expansion of the one-loop free energy,  near
the Hagedorn instability, is obtained.  The problem
of infrared regularization of thermodynamical quantities is pointed out. For
minimally coupling quantum fields related to the normal modes of
strings, the results are similar to the ones valid for Rindler space.
In the lower mass string states regime a connection with the quantum
corrections to the black hole entropy is outlined.
\end{abstract}

\vspace{8mm}


The behaviour  of  quantum  fields  in  backgrounds   with
horizons has been actively investigated during the last years
\cite{gibb78,birr82,thoo85,bomb86,sred93}.
In quantum field theory, the black hole entropy, calculated near
the  horizon, diverges \cite{thoo85}, and the interest for this long
standing problem has been growing
\cite{bomb86,sred93,barb94,dowk77,dowk78,dowk94,call94,suss94,barv94,frol94,deal94,bord94,cogn95,kaba95,oda95}.
The presence of the horizon involves  arbitrary  high  frequencies
and  it is argued that the string theory might be relevant for
the description of physics at ultra-short distances \cite{suss94,lowe94}.
Indeed the high temperature  behaviour  of  the  string  free  energy  can  be
extended in the presence of a black hole backgrounds
\cite{dabh94,dabho94,empar94}.
As a result of modular invariance, the closed string correction
to the entropy is ultraviolet finite.  The calculation of  the
quantum  correction  to  the  entropy of a large black hole in (non)-
critical string theory has  been also performed in
\cite{dabh94,eliz94,das95}.

The technical  difficulties in quantization are related to the
construction of the string spectrum in black hole backgrounds.
Conical  backgrounds  for strings,  interpreted in the form of
orbifolds \cite{dixo85}, have been applied to the black hole physics in
Ref. \cite{dabho94}.   Different   approaches   to   this  problem  give
approximately the same behaviour for the free energy (entropy)
\cite{dabho94,empar94}.

In this  paper,  we  shall  analyse  the  high  temperature
behaviour  for the open bosonic string free energy in hyperbolic
manifolds,  using a new representation (valid for different
tipes  of open and closed strings),  which has been introduced
in  Refs. \cite{byts92,byts93,bytse93,eliza94}. Our  approximation  will  be
carried  out in   space   of   vanishingly  small  curvature. For
massless quantum fields,  our techniques allows  to obtain
the  leading term of the high temperature expansion, which are  conformally
related to the one valid in Rindler space.

First of all, let us point out how the space-times with hyperbolic spatial
section may be  relevant in Rindler or black hole physics. To start with, we
shall consider the effective action
$\Ga [g,\beta]$
associated with a  conformal  scalar  field  in  D-dimensional
Rindler space,  which approximates the geometry outside a very
large black hole.  The metric of the Euclidean Rindler space can be  written
as follows
\beq
ds^2 = \xi^2 d \tau^2 + d\xi^2 + \sum_{i=1}^{N-1}dy_i^2 \, ,
\label{1}
\eeq
where $\tau$ is the Euclidean time (periodically  identified  with
period $\beta$), $y_i$ are  the  $(N-1)$-transverse  flat coordinates
$(D=N+1)$.  The  lines $\xi =\mbox{constant}$ correspond to observable
undergoing constant acceleration $\xi^{-1}$.

The one-loop effective action (up to  a  contribution  of  a  local
functional measure) can be presented in the form

\beq
\Ga [g,\beta] = \frac{1}{2} \Tr \ln A(\beta)\, ,
\label{2}
\eeq
where $A(\beta)\equiv \partial_{\tau}^2 + L_{N}$ , and $L_{N}$ is the standard
Laplace-Beltrami
operator acting in N-dimensional space. The entropy related to
the free energy ${\cal F}[g,\beta] \equiv \beta^{-1} \Ga [g,\beta]$
is
\beq
S = \beta^2 \frac{\partial}{\partial \beta} {\cal F}[g,\beta] =
\beta^2 \frac{\partial}{\partial \beta}
{\cal F} [g,\beta]_{Ren}\, ,
\label{3}
\eeq
where the  renormalization  of  the ultraviolet divergences of
the  free   energy   should   be  performed.  The   renormalized
free energies ${\cal F}[g,\beta ]_{Ren}$ for  two  conformally  related
static
spaces $g_{\mu \nu } = \exp{\at 2 \omega \ct} \bar{g}_{\mu \nu}$ are related as
follows

\beq
\lap {\cal F}[g,\omega] = {\cal F} [g,\beta]_{Ren} -
{\cal F}[\bar{g},\beta]\, .
\label{4}
\eeq
In four-dimensional case, the explicit form of the term
$\lap {\cal F} [g,\omega]$  can be found in Refs.
\cite{dowk88,dowk89,dowk90}.
However, the difference $\lap {\cal F} [g,\omega]$ for two conformally related
theories in a static spacetime, is
proportional to $\beta$ and  hence  does  not  contribute  to  the
entropy.
Further, we may apply Eq.(\ref{4}) to the  particular  case  of the Rindler
space. We have
$\omega = \fr{1}{2}\log{\xi^2}$,
and $\bar{g}_{\mu \nu} = \exp{\at -2\omega \ct}g_{\mu \nu}$ is the ultrastatic
metric.
As a result, conformally related fields in  the
manifold ${\cal M}$ of  the form ${\cal M} = S^1 \otimes H^N $, where $H^N$
is the simply connected real hyperbolic space, will be considered.
The free energy can be presented in the form
\beq
{\cal F}[\bar{g},\beta] = \fr{1}{2\beta}\Tr \ln {\bar{A}(\beta)}
= - \fr{1}{2\beta}\zeta'(0 |\bar{A}(\beta))\, ,
\label{6}
\eeq
where \cite{byts94}

\beqn
\zeta(s|\bar{A}(\be))& = &\frac{\beta \Ga(s-1/2)}{2\sqrt \pi \Ga(s)}\zeta
(s-\frac{1}{2} |\bar{L}_N)\nn \\
& + &\frac{1}{\sqrt \pi \Ga(s)2\pi i}\int _{\Re z =c}
dz\zeta_{R}(z)\Ga(\fr{z}{2})\Ga(\fr{z-1}{2}+s)\zeta(\fr{z-1}{2}+s |
\bar L_N)(\fr{\beta}{2})^{-z-1}\, ,
\label{7}
\eeqn
$c > N+1$,  $\zeta_{R}(z)$ is the Riemann zeta function, and the
self-adjoint Laplace-Beltrami operator $\bar {L}_N$ acting in the space
$H^N$. Finally, one can obtain (see Ref. \cite{byts94})

\beqn
{\cal F}[\bar g,\beta]_{Ren}& = &\frac{1}{2} \zeta(-\fr{1}{2}|
\bar{L}_N)_{Ren} - \frac{1}{2\pi i}\int_{\Re z=c}dz\zeta_{R}(z)\Ga(z-1)
\zeta(\fr{z-1}{2}|\bar{L}_N)\beta^{-z} \nn \\
&\equiv & F_0 + F(\beta)\, ,
\label{8}
\eeqn
with
\beq
\zeta(-\fr{1}{2}|\bar {L}_N)_{Ren} = FP \zeta (-\fr{1}{2}| \bar {L}_N)
+ (2-2\ln 2) Res_{z=-\fr{1}{2}}\zeta (z | \bar {L}_N)\, .
\label{9}
\eeq
Here $F_0$ is the vacuum energy, $F(\beta)$ is the temperature
dependent part of ${\cal F}[\bar g,\beta]_{Ren}$, the symbols FP
and $\Res$ denote the
finite  part and residue of the zeta function at the specified
point, respectively. Note that though zeta function diverges, it
does not depend on $\beta$ and its contribution to the entropy
vanishes.
For the massive quantum fields one can also use conformal transformation
techniques \cite{dowk88,dowk89,dowk90}
and construct a local zeta function $\ze(s,x|\bar{L}_N)$ (see for
example \cite{cogn95}).

Furthermore we shall be interested in massive quantum fields in even
dimensional manifolds,
i.e. $N$ is odd, with spatial hyperbolic sections $H^N$. Therefore the
Laplace-Beltrami operator $\bar{L}_N$ is acting in $H^N$. At the end,
we shall comment on the transition to the conformally invariant
massless case. The zeta function associated with operator $\bar {L}_N$
may be written in the form \cite{camp94,bytse92}

\beq
\zeta(z | \bar{L}_N) = \fr{1}{(4\pi)^{N/2}\Ga(N/2)}\sum_{k=1}
^{(N-1)/2}\alpha_{k,N}b^{1+2k-2z}B(k+\fr{1}{2}, z-k-\fr{1}{2})\, ,
\label{10}
\eeq
where $B(x,y)=\Ga(x)\Ga(y)/{\Ga(x+y)}$ is the Euler's beta  function,  the
coefficients $\alpha_{k,N}$ are defined by expanding the products into
polynomials in eigenvalues of $\bar{L}_N$ (see for detail
Refs.\cite{camp94,bytse92})
and $b$ is known constant depending on the mass of field and on
the curvature. Using the zeta function (\ref{10}), we obtain

\beqn
F(\beta) = &-& \frac{\sqrt \pi}{2\pi i(4\pi)^{\fr{N}{2}+1}\Ga(\fr{N}{2})}
\sum_{k=1}
^{\fr{N-1}{2}}\alpha_{k,N}\Ga(k+\fr{1}{2})\nn \\
&\times& \int_{\Re z=c} dz
\zeta_R(z)
\Ga(\fr{z}{2})\Ga(\fr{z}{2}-k-1)(\fr{\beta}{2})^{-z} b^{2+2k-z} \,.
\label{nm}
\eeqn

We may use the above formula, valid for the free energy of the quantum
fields,
in order to compute the one-loop free energy for the  bosonic string.
Strictly speaking, we should construct the string spectrum
in the manifold with non-vanishing curvature. For the flat manifold, the
spectrum, as an
infinite sum of quantum fields present  in normal modes of the string,
is well-known. To simplify the calculations, one
may assume that the curvature $R=-N(N-1)a^{-2}$ is vanishingly small
($a \rightarrow \infty$). In addition, we shall be interested in the
spectrum with almost linear Regge trajectories and the mass  formula, for
the open bosonic string, has the form
\cite{schw82,alva87,deve87,lars94}

\beq
M^2 = 2 \at \sum_{i=1}^{D-2}\sum_{n=1}^{\infty}
nN_{n}^{i}-1 \ct +O(a^{-4})\, ,
\label{12}
\eeq
where $N_{n}^{i}$ are the occupation numbers of the transverse
oscillators and the Regge slope parameter $\al'$ has been chosen equal to
one, for the sake of convenience.
For the minimally coupled scalar field of mass $m$,
$b^2=\rho_N^2+a^2m^2$, where $\rho_N=(N-1)/2$. In the limit $a \to
\ii$   the operator $b^2$, related to the quantum fields with mass
given by Eq. (\ref{12}), has the form $b^2=a^2M^2+O(1)$.
With the help
of heat kernel representation, the trace of the complex power of this operator
may be written in the form

\beq
\Tr b^{2k+2-z} = \frac{1}{\Ga(\fr{z}{2}-k-1)}\int_0^\ii dt
t^{\fr{z}{2}-k-2}\Tr \exp{\aq -t \at a^2 M^2 +O(1)\ct \cq }\,.
\label{13}
\eeq
Changing variables $t \rightarrow t \pi a^{-2}$ and performing
the trace over the entire Fock space we have

\beq
\Tr b^{2k+2-z} = \frac{1}{\Ga(\fr{z}{2}-k-1)}
(\frac{a^2}{\pi})
^{k+1-\fr{z}{2}}\int_0^\infty dt t^{\fr{z}{2}-k-2} \eta (it)^{-(N-1)}
\aq 1+O(a^{-2}) \cq\, ,
\label{14}
\eeq
where $\eta(\tau) = \exp{\at i\pi\tau/12 \ct}
\prod_{n=1}^{\infty}\aq 1-\exp{ \at -2\pi in\tau \ct}\cq $ is the Dedekind's
eta
function. Hence, for the free energy we obtain

\beqn
F(\beta) &=& -\frac{\sqrt{\pi}}{2\pi i(4\pi)^{N/2+1}\Ga(\fr{N}{2})}
\sum_{k=1}^{(N-1)/2}\al_{k,N}\Ga(k+\fr{1}{2}) \at \frac{a^2}{\pi}
\ct^{k+1}\nn \\
&\times& \int_{\Re z =c} dz \zeta_R(z) \Ga(\fr{z}{2})
\at \frac{\beta^2 a^2}{4\pi}\ct^{-\fr{z}{2}}
I_N(z) \aq 1+O(a^{-2}) \cq\, ,
\label{15}
\eeqn
where the integral $I_N(z)$ in Eq.(\ref{15}), namely
\beq
I_N(z)= \int_0^\ii dt t^{\fr{z}{2}-k-2}
\eta(it)^{-(N-1)}\,,
\eeq
should be regularized. In
the ultraviolet region, the integrand is not regular, since, for
$t\rightarrow 0$, one has
\beq
\eta(it)^{-(N-1)} = t^{(N-1)/2} e^{\pi (N-1)/12t} \aq 1+
O\at e^{-2\pi/t} \ct \cq \, .
\label{16}
\eeq
The divergence in the infrared region $(t \rightarrow \infty )$
is caused by
the tachyon in the string spectrum. The  tachyonic  divergence
signifies  a  vacuum  instability  which  may  be  important  to
understand its role in string theory. In order to regularize
the integral, it  is  more convenient  to  isolate the ultraviolet
divergence and introduce cutoff parameters

\beqn
 I_N(z)_{Reg} &=&
\int_0^\mu dt t^{z/2-k-2}
\at \eta(it)^{-(N-1)}-t^{(N-1)/2} \exp{\at \fr{\pi(N-1)}{12t}\ct} \ct \nn \\
&+&
\int_0^\ii dt t^{z/2-k-10} \exp{ \at -\fr{\pi(\epsilon-1)(N-1)}{12t}\ct}
\nn \\
& \equiv& G(\mu;\fr{z}{2}-k) + \aq
\frac{\pi}{12}(\epsilon-1)(N-1) \cq^{z/2-k+11}
\Ga(k-11-\fr{z}{2}) \, .
\label{17}
\eeqn
Here the regularization in the infrared region of the analytical part
$G(\mu;\fr{z}{2}-k)$ (the parameter $\mu$ provides the infrared
cutoff) has been done. On the next stage of calculations the ultraviolet
regularization will be removed $(\epsilon \to 0)$.

It is convenient to introduce the volume
$V_{D-2}=a^{D-2}V(\ca{F}_{D-2})$, where the constant factor
$V(\ca{F}_{D-2})$
can be referred to as the finite volume of the fundamental domain
associated with a compact hyperbolic manifold.
In the limit $a\to \infty$,
the leading term in the sum (\ref{15}) is the term with $k=(N-1)/2$,
and  we have (here we have restored the $\alpha'$ dependence)

\beqn
F(\beta)& =& -\frac{V_{D-2}a^2}{4\pi i V(\ca{F}_{D-2})}
(\frac{1}{4\pi^2 \al' })^{13}\int_{\Re z=c}
dz \zeta_{R}(z)\Ga(\fr{z}{2})\at \frac{\beta^2a^2}{4\pi \alpha'}\ct ^{-z/2}\nn
\\ &\times& \aq (-2\pi)^{z/2-1}\Ga(1-\fr{z}{2}) + G(\mu;\fr{z}{2}-12) \cq
 \aq 1+O(a^{-2}) \cq\,.
\label{18}
\eeqn

   The free  energy  (\ref{18})  can be present in terms of Laurent
series in inverse power of $\beta$  \cite{byts93}. In this approach, the
inverse critical Hagedorn  temperature $\beta_c = \pi \sqrt {8 \al'}$
arises, for the flat background, as
the convergence condition of the corresponding Laurent series.
The integrand (\ref{18}) is a meromorphic function which has first
order poles at $z=2n$, $n\in{\cal N}$.
    Moving the line of integration $\Re z=c$ to the left (and,
therefore, crossing the poles) one  can  obtain  for the  free
energy

\beq
F(\beta) = - \frac{\pi V_{D-2}}{12 (4\pi^2 \alpha')^{12} V(\ca{F}_{D-2})}
\beta^{-2} \aq 1+ \at 144 -\beta^{-2}\frac{4 \pi^4}{15}\ct
\frac{\al'}{a^{2}} \cq +O\at \at \frac{\al'}{a^{2}} \ct^2 \ct\,.
\label{19}
\eeq

We conclude with some remarks.
In this letter we have obtained the leading term in the high  temperature
expansion for the
free energy of open  bosonic  string. The corresponding entropy $S$
can be easily obtained using Eq. (\ref{3}). Our  analysis  have been
carried out for the open string in manifold $S^1 \otimes H^N$, with
vanishingly small curvature of the spatial section. By the way, the
asymptotic behaviour for the closed string free energy can be also
calculated with the help of the techniques presented here.
{}From  the  physical  point  of view, two  main types of
theories,  open  and  closed  strings,  admit   a similar high
temperature behaviour. For flat spaces ($a=\ii$), Eq. (\ref{19}) gives the
first term of the high temperature expansion of the free energy (see,
for example \cite{byts93}).

Furthermore, we would like to comment on the results we have obtained in
connection with the quantum corrections to the black hole entropy.
Let us recall that the energy density of a gas of free bosonic strings must be
small enough to be unafflicted by the Jeans
instability and large enough to contain many degrees of freedom, i.e.
$a^2 \gg \al'$. Near the Hagedorn temperature, the energy density behaves
as $ (\al')^{-13}$. Thus,
both the two conditions are satisfied if and only if
$G (\al')^{-12} \ll 1$ \cite{atic88,sanc90}, where the Newton constant $G$
defines the Plank length. For the lower mass states,
$  M^2 \propto (\al'/a^2)^2 $ \cite{deve87} and as a consequence,
the mass operator formally is vanishingly small. In this regime, our result
may be conformally related to the result obtained for massless quantum
fields in Rindler space or near black hole horizon.

The $\beta$-behaviour  of  the  free  energy  (\ref{19})  has  the
dependence  on  temperature  near  the  Hagedorn  transition
which  is similar to the one found in Ref. \cite{atic88}. Although the thermal
dependence in Eq. (\ref{19}), corresponds to quantum fields in two
dimensions, such a result can be interpreted as an indication of a
vast reduction of the fundamental degrees of freedom in string theory
\cite{atic88,empar94}. We point
out  that similar  leading  high temperature  expansion
has been obtained in \cite{dabho94,empar94}.  In fact, for the  free
energy  in  orbifold  string theory,  considered in Ref. \cite{dabho94}, the
large $N$ limit $(N=2\pi\beta^{-1})$ means that the one loop string
amplitude $A_N=-\beta F$
behaves like $\sim\beta^{-1}\log{\beta}$ and  therefore the leading
power $\beta$ -dependence is similar to the behaviour obtained
in our formalism.

\ack{A.A. Bytsenko thank I.N.F.N., Gruppo Collegato di Trento and the Physics
Department of Trento University for financial support and kind
hospitality.}


\begin{thebibliography}{10}


\bibitem{gibb78}
{ G.W. Gibbons and M.J. Perry}.
\newblock {Proc. R. Soc. Lond.}, {\bf {A358}}, {467}, (1978).

\bibitem{birr82}
{N.D. Birrell and P.C.  Davies}.
\newblock {Quantum Fields  in  Curved
    Space Time, Cambridge University Press}, (1982).

\bibitem{thoo85}
{ G. t' Hooft}.
\newblock {Nucl. Phys.}, {\bf {B256}}, {727}, (1985).

\bibitem{bomb86}
{L. Bombelli, R. Koul, J. Lee and R. Sorkin}.
\newblock {Phys. Rev.}, {\bf {D34}}, {373}, (1986).

\bibitem{sred93}
{M. Srednicki}.
\newblock {Phys. Rev. Lett.}, {\bf {71}}, {666}, (1993).

\bibitem{barb94}
{J.L.F. Barbon}.
\newblock {Phys. Rev.}, {\bf {D50}}, {2712}, (1994).

\bibitem{dowk77}
{J.S. Dowker}.
\newblock {J. Phys.}, {\bf {A10}}, {115}, (1977).

\bibitem{dowk78}
{J.S. Dowker}.
\newblock {Phys. Rev.}, {\bf {D18}}, {136}, (1978).

\bibitem{dowk94}
{J.S. Dowker}.
\newblock {Class. Quantum Grav.}, {\bf {11}}, {L55}, (1994).

\bibitem{call94}
{C. Callan and F. Wilczek}.
\newblock {Phys. Lett.}, {\bf {B333}}, {55}, (1994).

\bibitem{suss94}
{L. Susskind and U. Uglum}.
\newblock {Phys. Rev.}, {\bf {D50}}, {2700}, (1994).

\bibitem{barv94}
{A.O. Barvinsky, V.P. Frolov and A.I. Zelnikov}.
\newblock {Wavefunction of a Black Hole and the Dynamical Origin
of Entropy},  {Preprint }, {gr-qc/9404036}, (1994).

\bibitem{frol94}
{V.P. Frolov}.
\newblock {Why the Entropy of a Black Hole is A/4?}, {Preprint},
{gr-qc/9406037}, (1994).

\bibitem{deal94}
{S.P. de Alwis and N. Ohta}.
\newblock {On the Entropy of Quantum Fields in Black Hole Backgrounds},
{ Preprint}, {hep-th/9412027}, (1994).

\bibitem{bord94}
{M. Bordag and A.A. Bytsenko}.
\newblock {Quantum Corrections to the Entropy for Higher Spin Fields
in Hyperbolic Space},
{ Preprint}, {gr-qc/9412054}, (1994).

\bibitem{cogn95}
{G. Cognola, L. Vanzo and S. Zerbini}.
\newblock {One-Loop Quantum Corrections to the Free Energy for a
4-Dimensional Eternal Black Hole}, {Preprint U.T.F.}, {342},
(1995).

\bibitem{kaba95}
{D. Kabat}.
\newblock {Black Hole Entropy and Entropy of Entanglement},
{ Preprint}, {hep-th/9503016}, (1995).

\bibitem{oda95}
{I. Oda}.
\newblock {Thermodynamics of Black Hole in (N+3)-dimensions from
Euclidean N-brane Theory}, {Preprint}, {hep-th/9503009}, (1995).

\bibitem{lowe94}
{D.A. Lowe and A. Strominger}.
\newblock {Strings Near a Rindler or Black Hole Horizon}, {
Preprint}, {hep-th/9410215}, (1994).

\bibitem{dabh94}
{A. Dabholkar}.
\newblock {Quantum Corrections to Black Hole Entropy in String Theory},
{ Preprint}, {hep-th/9409158}, (1994).


\bibitem{dabho94}
{A. Dabholkar}.
\newblock {Strings on a Cone and Black Hole Entropy}, {Preprint},
{hep-th/9408098}, (1994).

\bibitem{empar94}
{R. Emparan}.
\newblock {Remarks on the Atick-Witten Behaviour and Strings Near
Black Hole Horizons}, {Preprint}, {hep-th/9412003}, (1994).

\bibitem{eliz94}
{E. Elizalde and S.D. Odintsov}.
\newblock {Non-Critical Bosonic String Corrections to the Black Hole
Entropy}, {Preprint}, {hep-th/9411024}, (1994).

\bibitem{das95}
{S.R. Das}.
\newblock {Geometric Entropy of Nonrelativistic Fermions and Two
Dimensional Strings}, {Preprint}, {hep-th/9501090}, (1995).

\bibitem{dixo85}
{L. Dixon, J. Harvey, C. Vafa and E. Witten}.
\newblock {Nucl. Phys.}, {\bf {B261}}, {678}, (1985).


\bibitem{byts92}
{A.A. Bytsenko, E. Elizalde, S.D. Odintsov and S. Zerbini}.
\newblock {Phys. Lett.}, {\bf {B297}}, {275}, (1992).

\bibitem{byts93}
{A.A. Bytsenko, E. Elizalde, S.D. Odintsov and S. Zerbini}.
\newblock {Nucl. Phys.}, {\bf {B394}}, {423}, (1993).

\bibitem{bytse93}
{A.A. Bytsenko, E. Elizalde, S.D. Odintsov and S. Zerbini}.
\newblock {Mod. Phys. Lett.}, {\bf {A8}}, {1131}, (1993).

\bibitem{eliza94}
{E. Elizalde, S.D. Odintsov, A. Romeo, A.A. Bytsenko and S. Zerbini}.
\newblock {Zeta Regularization Techniques with Applications}, {
{World Sci., Singapore}}, {1994}.

\bibitem{dowk88}
{J.S. Dowker and J.P. Schofield}.
\newblock {Phys. Rev.}, {\bf {D38}}, {3327}, (1988).

\bibitem{dowk89}
{J.S. Dowker}.
\newblock {Phys. Rev.}, {\bf {D39}}, {1235}, (1989).

\bibitem{dowk90}
{J.S. Dowker and J.P. Schofield}.
\newblock {J. Math. Phys.}, {\bf {31}}, {808}, (1990).

\bibitem{byts94}
{A.A. Bytsenko, G. Cognola, L. Vanzo and S. Zerbini}.
\newblock {Quantum Fields and Extended Objects in Space Times with
Constant Curvature Spatial Section}, {Preprint U.T.F.}, {325},
(1994).

\bibitem{camp94}
{R. Camporesi and A. Higuchi}.
\newblock {J. Math. Phys.}, {\bf {35}}, {4217}, (1994).


\bibitem{bytse92}
{A.A. Bytsenko, L. Vanzo and S. Zerbini}.
\newblock {Mod. Phys. Lett.}, {\bf {A7}}, {397}, (1992).

\bibitem{schw82}
{J. Schwarz}.
\newblock {Phys. Rep.}, {\bf {89}}, {223}, (1982).

\bibitem{alva87}
{E. Alvarez and M.A.R. Osorio}.
\newblock {Phys. Rev.}, {\bf {D36}}, {1175}, (1987).

\bibitem{deve87}
{H.J. De Vega and N. Sanchez}.
\newblock {Phys. Lett.}, {\bf {B197}}, {320}, (1987).

\bibitem{lars94}
{A.L. Larsen and N. Sanchez}.
\newblock {Mass Spectrum of Strings in Anti de Sitter Spacetime},
{Preprint}, {hep-th/9410132}, (1994).

\bibitem{atic88}
{J.J. Atick and E. Witten}.
\newblock {Nucl. Phys.}, {\bf {B310}}, {291}, (1988).

\bibitem{sanc90}
{N. Sanchez and G. Veneziano}.
\newblock {Nucl. Phys.}, {\bf {B333}}, {253}, (1990).


\end{thebibliography}
\end{document}